# Mathematical proof concerning the additivity problem of nonlinear normalized citation counts

Xing Wang[a, b, *], Zhihui Zhang[c]

[a]School of Information, Shanxi University of Finance & Economics, 030006 Taiyuan, China
[b]Shanxi Key Laboratory of Data Element Innovation and Economic Decision Analysis, 030006 Taiyuan, China
[c]Tongfang Knowledge Network Technology Co., Ltd., 100192 Beijing, China

[*]Corresponding author. Email: wangxing@sjtu.edu.cn

**Abstract** The issue of whether nonlinear normalized citation counts can be added is critically important in scientometrics because it touches upon the theoretical foundation of underlying computation in the field. In this paper, we provide rigorous mathematical proofs for the key theorems underlying this fundamental issue. Based on these proofs, we ultimately arrive at the following conclusion: a nonlinear normalization method for citation counts must be a non-equidistant transformation; consequently, the resulting nonlinear normalized citation counts are no longer equidistant and therefore cannot be added. Furthermore, because our mathematical proofs are established over the real number domain, we also derive a more general conclusion that is applicable to data transformations over the real number domain across various scientific fields: a nonlinear transformation becomes a non-equidistant transformation only if it satisfies a certain regularity condition, for example, if this nonlinear transformation is a continuous function, monotonic over a certain interval, or its domain is restricted to rational numbers. In such cases, the resulting nonlinear data are no longer equidistant and therefore cannot be added. This general conclusion can be broadly applied to various linear and nonlinear transformation problems, which offers significant insights for addressing the misuse of nonlinear data.

**Keywords** normalized citation counts; nonlinear normalization method; nonlinear transformation; linear transformation; equidistant transformation; non-equidistant transformation

# 1. Introduction

In the evaluation of scientific research impact based on citation analysis, raw citation counts of papers from different fields cannot be compared directly because of inherent differences in citation practices among fields. Similarly, raw citation counts vary by publication year and document type, rendering direct comparisons across these dimensions equally problematic. In such cases, to ensure a fair evaluation of the scientific impact of papers across different fields, publication years, and document types, it is necessary to use citation normalization methods to minimize or eliminate disciplinary, temporal, and document-type differences. The normalized citation counts obtained after using normalization methods are independent of the field, publication time, and document type of the paper so that the comparison of citation impact across different fields, publication years, and document types can be naturally achieved through the comparison of normalized citation counts. As such, normalized citation counts have become one of the most widely used indicators in research evaluation.

To date, numerous normalization methods have been proposed in the academic community (Bornmann, 2020; Gómez-Déniz & Dorta-González, 2024; Tong et al., 2023; Thelwall, 2017; Waltman, 2016; Wang, 2024). In our recent paper published in the *Journal of Informetrics* (Wang, 2024), we classified the existing main normalization methods into linear normalization methods (e.g., mean-based method and z-score method) and nonlinear normalization methods (e.g., percentile rank method and logarithmic transformation method), and proved why nonlinear normalized citation counts obtained using nonlinear normalization methods cannot be added from the perspective of theoretical analysis in mathematics.

The fundamental reason that nonlinear normalized citation counts cannot be added is that the procedure of implementing a nonlinear normalization transformation is a non-equidistant transformation; consequently, each citation count of the nonlinear normalized citation counts obtained through such a transformation is no longer equidistant. However, a prerequisite for data that can be used for arithmetic operations (e.g., addition) is that the measuring unit of the data (in the context of this paper, each citation count) is equidistant. Therefore, these nonlinear normalized citation counts cannot be added.

In our previous paper (Wang, 2024), we proved that the nonlinear transformation must be a non-equidistant transformation, which thus proved why nonlinear normalized citation counts cannot be added. However, the proof method in our previous paper was more of an intuitive and easy to understand proof method combined with graphics, which may not be mathematically rigorous. In this paper, we provide a more mathematically rigorous proof method to provide more complete and stronger theoretical support for the conclusion that nonlinear normalized citation counts cannot be added.

In fact, the question of whether nonlinear normalized citation counts can be added has sparked debate within the scientometric community (Donner, 2022; Haunschild & Bornmann, 2025; Wang, 2024, 2025; Zhang et al., 2015). Some scholars do not believe that nonlinear normalized citations are definitely not additive (Haunschild & Bornmann, 2025). Moreover, there are also many examples in academic and evaluation practice circles where nonlinear normalized citations have been added. For instance, the well-known citation database provider Clarivate offers indicators such as "Average Percentile" and "Average JIF Percentile" in its InCites database, which involve the addition of nonlinear normalized citation counts (Clarivate, 2025). Similarly, some academic papers use indicators such as the "Mean Normalized Log Citation Score (MNLCS)," which also involve the addition of nonlinear normalized citation counts (Thelwall, 2025; Thelwall et al., 2023, 2024). Given both the theoretical debates and widespread practical misuse of nonlinear normalized citation indicators, this paper seeks to bring this controversial issue to a definitive conclusion through rigorous mathematical proofs.

We present the mathematical proofs and corresponding analyses in this paper step by step. We begin by proposing the following four theorems that are closely related to the question of whether nonlinear normalized citation counts can be added. These four theorems form the foundation for all subsequent analysis in this study.

**Theorem 1:** A linear transformation must be an equidistant transformation.
**Theorem 2:** If a continuous function $f(x)$ is an equidistant transformation, then it must be a linear transformation.
**Theorem 3:** A non-equidistant transformation must be a nonlinear transformation.
**Theorem 4:** If a continuous function $f(x)$ is a nonlinear transformation, then it must be a non-equidistant transformation.

We prove Theorems 1 and 2 in Sections 2 and 3, respectively, and Theorems 3 and 4 in Section 4. With these four theorems established as the theoretical groundwork, we then proceed to the Discussion section, where we conduct a more in-depth analysis of the additivity problem of nonlinear normalized citation counts.

Based on the proofs of the four theorems and the in-depth analysis in the Discussion section, we finally arrive at a mathematically rigorous conclusion regarding the issue of "whether nonlinear normalized citation counts can be added": a nonlinear normalization method for citation counts must be a non-equidistant transformation; as a result, the nonlinear normalized citation counts obtained through such a transformation are no longer equidistant and therefore cannot be added.

Furthermore, it should be particularly emphasized that we establish all the mathematical proofs in this paper over the entire real number domain to ensure their generality and broad applicability. Consequently, the results of these proofs are universal and can be broadly applied to various linear/nonlinear and equidistant/non-equidistant transformation problems over the real number domain across various scientific fields. Accordingly, we also obtain more general conclusions concerning data transformation problems. These conclusions offer significant insights for addressing the misuse of nonlinear data in a range of scientific disciplines. We present these conclusions in the Conclusion section, where we also summarize our study and offer related reflections.

## 2. Proof of Theorem 1

In this section, we provide the proof of Theorem 1.

**Theorem 1.** A linear transformation must be an equidistant transformation.

**Proof.** Let us first define an arbitrary linear transformation of the form

$$y = f(x) = kx + b,$$

where the domain and codomain of the function are both the set of real numbers, i.e., $f : \mathbb{R} \to \mathbb{R}$. Here, $x$ denotes the value before the transformation (in the context of this paper, it refers to the raw citation count of one paper), $y$ denotes the value after the transformation (in the context of this paper, it refers to the normalized citation count of this paper), and $k \neq 0$ is a constant scaling factor, and $b$ is a constant shift.

Then, let the independent variable $x$ take four arbitrary real values $x_1, x_2, x_3, x_4 \in \mathbb{R}$, such that

$$x_1 - x_2 = x_3 - x_4. \tag{1}$$

To prove Theorem 1 (i.e., a linear transformation must be an equidistant transformation), based on the definition of the equidistant transformation (Wang, 2024), we need to show that the corresponding transformed values $y_1, y_2, y_3, y_4$ satisfy

$$y_1 - y_2 = y_3 - y_4, \tag{2}$$

where $y_i = f(x_i) = kx_i + b$ for $i$ = 1, 2, 3, 4.

The following proof is straightforward.

Given the linear transformation $y = f(x) = kx + b$ with $k \neq 0$, the outputs corresponding to $x_1, x_2, x_3, x_4$ are:

$$y_1 = f(x_1) = kx_1 + b,$$

$$y_2 = f(x_2) = kx_2 + b,$$

$$y_3 = f(x_3) = kx_3 + b,$$

$$y_4 = f(x_4) = kx_4 + b.$$

Next, we calculate the differences $y_1 - y_2$ and $y_3 - y_4$:

$$y_1 - y_2 = (kx_1 + b) - (kx_2 + b) = k(x_1 - x_2),$$

$$y_3 - y_4 = (kx_3 + b) - (kx_4 + b) = k(x_3 - x_4).$$

According to Eq. (1) $x_1 - x_2 = x_3 - x_4$, it follows that

$$y_1 - y_2 = k(x_1 - x_2) = k(x_3 - x_4) = y_3 - y_4.$$

Hence, Eq. (2) $y_1 - y_2 = y_3 - y_4$ is established, which confirms that the differences between the transformed values are equal whenever the differences between their corresponding input values are equal. This satisfies the definition of an equidistant transformation (Wang, 2024). Therefore, we have proved Theorem 1: a linear transformation must be an equidistant transformation. **Q.E.D.**

## 3. Proof of Theorem 2

In this section, we provide the proof of Theorem 2.

**Theorem 2.** If a continuous function $f(x)$ is an equidistant transformation, then it must be a linear transformation.

**Proof.** Let $f : \mathbb{R} \to \mathbb{R}$ be a continuous function satisfying the following property: for any $x_1, x_2, x_3, x_4 \in \mathbb{R}$, if

$$x_1 - x_2 = x_3 - x_4,$$

then

$$f(x_1) - f(x_2) = f(x_3) - f(x_4).$$

This property defines $y = f(x)$ as an equidistant transformation, where $x$ denotes the value before the transformation (in the context of this paper, it refers to the raw citation count of one paper) and $y$ denotes the value after the transformation (in the context of this paper, it refers to the normalized citation count of this paper).

We now aim to prove that any continuous function $f(x)$ satisfying this property must be a linear transformation, that is,

$$f(x) = kx + b$$

for constants $k, b \in \mathbb{R}$.

We divide our proof into three steps.

**Step 1: We define an auxiliary function $g(x)$ and show that it also satisfies the equidistant transformation property.**

Let us define an auxiliary function as follows:

$$g(x) = f(x) - f(0). \tag{3}$$

Because the function $f(x)$ is continuous, the function $g(x) = f(x) - f(0)$ is also continuous. Furthermore, we have

$$g(0) = f(0) - f(0) = 0. \tag{4}$$

Given that $f(x)$ is an equidistant transformation, for any real numbers $x_1, x_2, x_3, x_4$ satisfying

$$x_1 - x_2 = x_3 - x_4,$$

it follows that

$$f(x_1) - f(x_2) = f(x_3) - f(x_4). \tag{5}$$

Now we calculate the differences $g(x_1) - g(x_2)$ and $g(x_3) - g(x_4)$:

$$g(x_1) - g(x_2) = [f(x_1) - f(0)] - [f(x_2) - f(0)] = f(x_1) - f(x_2),$$

$$g(x_3) - g(x_4) = [f(x_3) - f(0)] - [f(x_4) - f(0)] = f(x_3) - f(x_4).$$

Hence, according to Eq. (5) $f(x_1) - f(x_2) = f(x_3) - f(x_4)$, we have

$$g(x_1) - g(x_2) = g(x_3) - g(x_4);$$

that is, for any $x_1, x_2, x_3, x_4 \in \mathbb{R}$, if

$$x_1 - x_2 = x_3 - x_4,$$

then

$$g(x_1) - g(x_2) = g(x_3) - g(x_4),$$

which shows that the function $g(x)$ also satisfies the equidistant transformation property.

**Step 2: We prove that the function $g(x)$ is an additive function, that is, it satisfies the Cauchy functional equation.**

For any $a, b \in \mathbb{R}$, consider the following:

$$(a+b) - b = a = a - 0.$$

Because $g(x)$ satisfies the equidistant transformation property (established in Step 1), it follows that:

$$g(a+b) - g(b) = g(a) - g(0). \tag{6}$$

Moreover, because $g(0) = 0$ (as shown in Eq. (4)), Eq. (6) simplifies to

$$g(a+b) - g(b) = g(a).$$

Rearranging terms yields:

$$g(a+b) = g(a) + g(b). \tag{7}$$

Therefore, for all $a, b \in \mathbb{R}$, we have $g(a+b) = g(a) + g(b)$, which proves that $g(x)$ is an additive function on $\mathbb{R}$, that is, it satisfies the Cauchy functional equation.

**Step 3: We deduce that $f(x)$ is a linear transformation from the fact that $g(x)$ is a continuous additive function on $\mathbb{R}$.**

Because the function $g(x)$ is both continuous (established in Step 1) and additive on $\mathbb{R}$ (i.e., it satisfies the Cauchy functional equation, as proven in Step 2), it follows from the classical result on the Cauchy functional equation that

$$g(x) = kx \tag{8}$$

for some constant $k \in \mathbb{R}$.

This result on Cauchy functional equation is well-established in the literature (see, e.g., Aczél, 1966; Reem, 2017; Small, 2007).

Recalling the definition of $g(x)$ from Eq. (3),

$$g(x) = f(x) - f(0),$$

we substitute Eq. (8) into Eq. (3) to obtain:

$$f(x) = g(x) + f(0) = kx + f(0).$$

Letting $b = f(0)$, for any $x \in \mathbb{R}$, we have

$$f(x) = kx + b,$$

which confirms that $f(x)$ must be a linear transformation and thus completes the proof of Theorem 2. **Q.E.D.**

## 4. Proofs of Theorem 3 and 4

In this section, we provide the proofs of Theorems 3 and 4.

**Theorem 3:** A non-equidistant transformation must be a nonlinear transformation.

**Proof.** Because we have already proved Theorem 1, that is, a linear transformation must be an equidistant transformation, the contrapositive of Theorem 1, that is, a non-equidistant transformation must be a nonlinear transformation, must also hold. This completes the proof of Theorem 3. **Q.E.D.**

**Theorem 4:** If a continuous function $f(x)$ is a nonlinear transformation, then it must be a non-equidistant transformation.

**Proof.** The proof of Theorem 4 follows a similar line of reasoning to that of Theorem 3. Because we have already proved Theorem 2, that is, if a continuous function $f(x)$ is an equidistant transformation, then it must be a linear transformation, the contrapositive of Theorem 2, that is, if a continuous function $f(x)$ is a nonlinear transformation, then it must be a non-equidistant transformation, must also be true. This completes the proof of Theorem 4. **Q.E.D.**

Thus far, we have completed the proofs of Theorems 1–4.

## 5. Discussion

### *5.1 Theoretical foundations: the Cauchy functional equation and related regularity conditions*

An interesting and worthwhile issue for further discussion is the following: Compared with our previous study (Wang, 2024), this study requires the imposition of a restriction condition that the nonlinear transformation must be a continuous function. Only under this condition of a "continuous function" can we conclude that "this nonlinear transformation method is a non-equidistant transformation, and the nonlinear data obtained through this nonlinear transformation method are no longer equidistant, therefore, these nonlinear data cannot be added."

Without the "continuous function" precondition, the proof of Theorem 2 would fail to derive the solution *g*(*x*) = *kx* for the additive function *g*(*x*). As a result, we could not establish the proposition that "an equidistant transformation must be a linear transformation." Furthermore, this would prevent us from establishing the validity of its contrapositive: "a nonlinear transformation must be a non-equidistant transformation" (i.e., Theorem 4). Ultimately, the conclusion that "nonlinear data cannot be added" would be unattainable.

Certainly, if we do not impose the restriction condition of a "continuous function" on the nonlinear transformation method, but instead impose some other restriction conditions, such as requiring the nonlinear transformation method to be monotonic on a certain interval, or to be bounded above or below on a certain interval, then such a nonlinear transformation method is still a non-equidistant transformation. Consequently, the nonlinear data obtained through this transformation are also no longer equidistant and therefore cannot be added. In the following, we provide a detailed explanation of this issue.

Before explaining this issue in detail, we need first to lay some groundwork regarding the Cauchy functional equation. Let us revisit Eqs. (3), (7), and (8) from Section 3. In the proof of Theorem 2, we constructed an auxiliary function *g*(*x*), as defined in Eq. (3). Eq. (7) shows that the function *g*(*x*) is an additive function, that is, it satisfies the Cauchy functional equation. According to the established theory concerning the Cauchy functional equation (see, e.g., Aczél, 1966; Reem, 2017; Small, 2007), if we wish to solve this functional equation over the set of real numbers to obtain a "nice" solution like Eq. (8), *g*(*x*) = *kx*, then we need to impose certain "regularity conditions" on *g*(*x*), such as the following:

- Condition 1: the function *g*(*x*) is a continuous function.
- Condition 2: the function *g*(*x*) is monotonic on a certain interval.
- Condition 3: the function *g*(*x*) is bounded above or below on a certain interval.
- Condition 4: the function *g*(*x*) is Lebesgue measurable.
…

Only if *g*(*x*) satisfies at least one of these "regularity conditions" can we obtain the "nice" solution of the form *g*(*x*) = *kx*. Otherwise, if we do not impose any regularity conditions on the function *g*(*x*), then the function *g*(*x*) may yield many so-called "pathological solutions." Notably, if the domain of *g*(*x*) is restricted to the set of rational numbers ℚ, then the solution is unique and also of the form *g*(*x*) = *kx*.

At this point, our theoretical groundwork concerning the Cauchy functional equation is complete. In Section 3, the proof of Theorem 2 relied on Condition 1. In the following section, we demonstrate Conditions 2 and 3 also suffice to establish the non-additivity of the nonlinear data.

## 5.2 When the nonlinear transformation method satisfies the regularity condition of monotonicity

We now prove that if a nonlinear transformation is monotonic on a certain interval, then it must be a non-equidistant transformation; as a result, the nonlinear data obtained from such a transformation are no longer equidistant and thus cannot be added. To establish this, we need to demonstrate that the following Theorem 5 and its contrapositive, Theorem 6, hold true.

**Theorem 5:** Let $f(x)$ be a function that is monotonic on a certain interval $(c, d)$, where $c, d \in \mathbb{R}$ and $c < d$. If $f(x)$ is an equidistant transformation, then it must be a linear transformation.

**Theorem 6:** Let $f(x)$ be a function that is monotonic on a certain interval $(c, d)$, where $c, d \in \mathbb{R}$ and $c < d$. If $f(x)$ is a nonlinear transformation, then it must be a non-equidistant transformation.

**Proof of Theorem 5.** Let $f: \mathbb{R} \to \mathbb{R}$ be a function satisfying the conditions of Theorem 5, that is, it is monotonic on a certain interval $(c, d)$ and is an equidistant transformation. The proof of Theorem 5 follows exactly the same procedure as that of Theorem 2. We still need to construct an auxiliary function $g(x)$:

$$g(x) = f(x) - f(0).$$

Given that $f(x)$ is monotonic on the interval $(c, d)$, the auxiliary function $g(x) = f(x) - f(0)$ is also monotonic on $(c, d)$. Moreover, we can readily show that $g(x)$ satisfies the Cauchy functional equation; that is, for any $a, b \in \mathbb{R}$,

$$g(a+b) = g(a) + g(b),$$

and the proof of this property follows the same reasoning as that in the proof of Theorem 2.

Because $g(x)$ is monotonic on $(c, d)$, by the theory of the Cauchy functional equation discussed earlier, we may apply Condition 2 (monotonicity regularity) to deduce that $g(x) = kx$ for some constant $k$. Substituting $g(x) = kx$ back into the expression $g(x) = f(x) - f(0)$ yields:

$$f(x) = kx + b \quad \text{where} \quad b = f(0),$$

which confirms that $f(x)$ is a linear transformation. This completes the proof of Theorem 5.                                                                                                          **Q.E.D.**

**Proof of Theorem 6.** Theorem 6 is the contrapositive of Theorem 5. Therefore, it holds by logical equivalence.                                                                                      **Q.E.D.**

At this point, based on Theorem 6, we can conclude the following: If a nonlinear transformation is monotonic on a certain interval, then it must be a non-equidistant transformation; as a result, the nonlinear data obtained from such a transformation are no longer equidistant and thus cannot be added.

### *5.3 When the nonlinear transformation method satisfies the regularity condition of boundedness*

We now extend our analysis by demonstrating that if a nonlinear transformation is bounded above or below on a certain interval, then it must be a non-equidistant transformation; as a result, the nonlinear data obtained from such a transformation are no longer equidistant and thus cannot be added.

Similarly, to establish the validity of this conclusion, we introduce Theorems 7 and 8, which parallel Theorems 5 and 6 but invoke boundedness rather than monotonicity as the regularity condition.

**Theorem 7:** Let $f(x)$ be a function that is bounded above (or below) on a certain interval $(c, d)$, where $c, d \in \mathbb{R}$ and $c < d$. If $f(x)$ is an equidistant transformation, then it must be a linear transformation.

**Theorem 8:** Let $f(x)$ be a function that is bounded above (or below) on a certain interval $(c, d)$, where $c, d \in \mathbb{R}$ and $c < d$. If $f(x)$ is a nonlinear transformation, then it must be a non-equidistant transformation.

**Proof of Theorem 7.** Let $f : \mathbb{R} \to \mathbb{R}$ be a function satisfying the conditions of Theorem 7. As in previous proofs, we define the auxiliary function $g(x) = f(x) - f(0)$.

Given that $f(x)$ is bounded above (or below) on the interval $(c, d)$, the auxiliary function $g(x) = f(x) - f(0)$ is also bounded above (or below) on the interval $(c, d)$. Moreover, we can readily show that $g(x)$ satisfies the Cauchy functional equation. The proof of this property follows the same reasoning as that of Theorem 2.

Because $g(x)$ is bounded above (or below) on the interval $(c, d)$, we invoke Condition

3 (boundedness regularity) to deduce that $g(x) = kx$ for some constant $k$. Substituting this result back into $g(x) = f(x) - f(0)$ yields:

$$f(x) = kx + b \text{ where } b = f(0),$$

which confirms that $f(x)$ is a linear transformation. This completes the proof of Theorem 7. **Q.E.D.**

**Proof of Theorem 8.** Theorem 8 is the contrapositive of Theorem 7 and therefore holds by logical equivalence. **Q.E.D.**

Therefore, based on Theorem 8, we can conclude the following: If a nonlinear transformation is bounded above or below on a certain interval, then it must be a non-equidistant transformation; as a result, the nonlinear data obtained from such a transformation are no longer equidistant and thus cannot be added.

At this point, we can clearly see that if we want to obtain the conclusion that "a nonlinear transformation is a non-equidistant transformation, and the nonlinear data obtained from such a transformation are no longer equidistant" in the real number domain, we must impose certain "regularity conditions" on the nonlinear transformation. Examples of such conditions include requiring the nonlinear transformation to be a continuous function, monotonic on a certain interval, or bounded above or below on a certain interval.

*5.4 When the domain of the nonlinear transformation method is the set of rational numbers*

We now return to the practical context of this study—namely, the normalization of citation counts. Because citation counts are all positive integers or 0, they all naturally fall within the set of rational numbers $\mathbb{Q}$. This observation has important implications for the theoretical treatment of normalization methods.

Whether the transformation method is linear or nonlinear, its domain is restricted to $\mathbb{Q}$. With this in mind, we introduce the following two theorems to address the additivity of nonlinear normalized citation counts.

**Theorem 9:** Let $f : \mathbb{Q} \to \mathbb{R}$ be a function. If $f(x)$ is an equidistant transformation, then it must be a linear transformation.

**Theorem 10:** Let $f : \mathbb{Q} \to \mathbb{R}$ be a function. If $f(x)$ is a nonlinear transformation, then it must be a non-equidistant transformation.

**Proof of Theorem 9.** Let $f: \mathbb{Q} \to \mathbb{R}$ be an equidistant transformation. As in previous proofs, we define the auxiliary function $g(x) = f(x) - f(0)$.

Given that the domain of $f(x)$ is the set of rational numbers, the domain of the auxiliary function $g(x) = f(x) - f(0)$ is also the set of rational numbers. Moreover, it is straightforward for us to show that $g(x)$ satisfies the Cauchy functional equation, with the proof proceeding in the same way as that for Theorem 2.

Because the domain of $g(x)$ is the set of rational numbers, we can invoke the established results on the Cauchy functional equation over the rational domains to conclude that $g(x) = kx$ for some constant $k$. Substituting back, we obtain:

$$f(x) = kx + b \text{ where } b = f(0),$$

which confirms that $f(x)$ is a linear transformation. This completes the proof of Theorem 9. **Q.E.D.**

**Proof of Theorem 10**. Theorem 10 is the contrapositive of Theorem 9, and therefore holds by logical equivalence. **Q.E.D.**

Therefore, based on Theorem 10, we can conclude the following: If the domain of a nonlinear transformation is rational numbers, then this nonlinear transformation method must be a non-equidistant transformation; as a result, the nonlinear data obtained from such a transformation are no longer equidistant and thus cannot be added.

This result is particularly significant for citation normalization. Because the domain of any nonlinear normalization method must lie within the set of rational numbers, any nonlinear normalization method must necessarily be a non-equidistant transformation, without requiring additional regularity conditions. Consequently, the nonlinear normalized citation counts obtained from such a transformation are no longer equidistant, and therefore, they cannot be added. This brings a definitive conclusion to the question of whether nonlinear normalized citation counts can be added.

## 6. Conclusion

In this paper, we began by posing the question of whether nonlinear normalized citation counts can be added and proposed four fundamental mathematical theorems underlying this question. We provided rigorous proofs for each of them. Based on these proofs and the extended analysis presented in the Discussion section, we finally draw the following three key conclusions, thereby answering the original question:

First, for data transformation problems over the real number domains, a nonlinear transformation becomes a non-equidistant transformation only if it satisfies a certain regularity condition, for example, if this nonlinear transformation is a continuous function, or if it is monotonic on a certain interval, or if it is bounded above (or below) on a certain interval. In such cases, the resulting nonlinear data are no longer equidistant and therefore cannot be added.

Second, for data transformation problems where the domain is restricted to rational numbers, a nonlinear transformation must be a non-equidistant transformation. Therefore, the nonlinear data obtained through such a transformation are no longer equidistant and thus cannot be added.

Third, a nonlinear normalization method for citation counts must be a non-equidistant transformation; as a result, the nonlinear normalized citation counts obtained through such a transformation are no longer equidistant and therefore cannot be added.

In this paper, we provide a definitive answer to the central question of whether nonlinear normalized citation counts can be added: they cannot. This question is of critical importance in scientometrics. On the one hand, adding normalized citation counts has a very wide range of application scenarios (e.g., actual research evaluation is often conducted at aggregation levels, such as at the level of institutions, journals and countries, which requires adding the normalized citation counts of individual papers to measure the overall or average research impact of the evaluated entity at the aggregation level). On the other hand, both academia and evaluation practice circles frequently misuse nonlinear normalization methods by adding nonlinear normalized citation counts, despite their non-equidistant nature. It can be said that the question of "whether nonlinear normalized citation counts can be added" touches upon the theoretical foundation of the underlying computation of scientometrics. If this foundation is not clearly defined, many computations in the field of scientometrics would be meaningless.

The mathematical proofs presented in this paper provide a solid and comprehensive theoretical foundation for the conclusion that nonlinear normalized citation counts cannot be added, which can help guide academia and evaluation practice circles toward a more scientific and rational use of citation normalization methods in the future, and thereby help to avoid the continued misuse of nonlinear normalization methods.

Moreover, beyond the field of scientometrics, there are also numerous cases in other scientific fields in which nonlinear data are improperly added, resulting in the misuse of nonlinear transformation methods (e.g., Anderegg et al., 2022; Brooks et al., 2024; Caldwell et al., 2024; Marquardt et al., 2022; Parks et al., 2023; Zhang & Gosline, 2023; Zheng et al., 2022). Because the mathematical proofs in this paper are established over the entire domain of real numbers, the general conclusions derived here are broadly applicable and offer valuable insights for addressing the misuse of nonlinear transformation methods in other scientific fields.

# Acknowledgment

This research was supported by MOE (Ministry of Education in China) Project of Humanities and Social Science (Project No. 22YJCZH180).

16Thelwall, M. (2025). ChatGPT ranking of business and management journals with article quality scores. *Aslib Journal of Information Management*, advance online publication. https://eprints.whiterose.ac.uk/id/eprint/225440/ (Published online 13 May 2025)

Thelwall, M., Kousha, K., Abdoli, M., Stuart, E., Makita, M., Wilson, P., & Levitt, J. (2023). Why are co-authored academic articles more cited: Higher quality or larger audience? *Journal of the Association for Information Science and Technology*, 74(7), 791–810.

Thelwall, M., Kousha, K., Abdoli, M., Stuart, E., Makita, M., Wilson, P., & Levitt, J. (2024). Which international co-authorships produce higher quality journal articles? *Journal of the Association for Information Science and Technology*, 75(7), 769–788.

Tong, S., Chen, F., Yang, L., & Shen, Z. (2023). Novel utilization of a paper-level classification system for the evaluation of journal impact: An update of the CAS Journal Ranking. *Quantitative Science Studies, 4*(4), 960–975.

Waltman, L. (2016). A review of the literature on citation impact indicators. *Journal of Informetrics, 10*(2), 365–391.

Wang, X. (2024). The misuse of the nonlinear field normalization method: Nonlinear field normalization citation counts at the paper level should not be added or averaged. *Journal of Informetrics, 18*(3), 101531.

Wang, X. (2025). Response to "Conclusions need to follow from supporting results" by Haunschild and Bornmann. *Journal of Informetrics, 19*(2), 101666.

Zhang, Y., & Gosline, R. (2023). Human favoritism, not AI aversion: People's perceptions (and bias) toward generative AI, human experts, and human-GAI collaboration in persuasive content generation. *Judgment and Decision Making, 18*, e41.

Zhang, Z., Cheng, Y., & Liu, N. C. (2015). Improving the normalization effect of mean-based method from the perspective of optimization: optimization-based linear methods and their performance. *Scientometrics, 102*(1), 587–607.

Zheng, H., Goh, D. H., Lee, E. W. J., Lee, C. S., & Theng, Y. (2022). Understanding the effects of message cues on COVID-19 information sharing on Twitter. *Journal of the Association for Information Science and Technology, 73*(6), 847–862.